\documentclass[twocolumn]{aastex63}
\usepackage{graphicx}

\newcommand\gildas{\textsc{gildas}}

\received{XXX}
\revised{YYY}
\accepted{ZZZ}
\submitjournal{ApJ}

\shorttitle{Redshifts of the Brightest GOODS-N Submillimeter Galaxies}
\shortauthors{Jones et al.}
\graphicspath{{./}{figures/}}
\begin{document}

\title{NOEMA Redshift Measurements of Extremely Bright Submillimeter Galaxies Near the GOODS-N}

\correspondingauthor{Logan Jones}
\email{ljones@astro.wisc.edu}

\author[0000-0002-1706-7370]{Logan H. Jones}
\affiliation{Department of Astronomy, University of Wisconsin-Madison, 475 N. Charter Street, Madison, WI 53706, USA}

\author[0000-0003-3910-6446]{Michael J. Rosenthal}
\affiliation{Department of Astronomy, University of Wisconsin-Madison, 475 N. Charter Street, Madison, WI 53706, USA}

\author[0000-0002-3306-1606]{Amy J. Barger}
\affiliation{Department of Astronomy, University of Wisconsin-Madison, 475 N. Charter Street, Madison, WI 53706, USA}
\affiliation{Department of Physics and Astronomy, University of Hawaii,
2505 Correa Road, Honolulu, HI 96822, USA}
\affiliation{Institute for Astronomy, University of Hawaii, 2680 Woodlawn Drive,
Honolulu, HI 96822, USA}

\author[0000-0002-6319-1575]{Lennox L. Cowie}
\affiliation{Institute for Astronomy, University of Hawaii, 2680 Woodlawn Drive,
Honolulu, HI 96822, USA}



%TC:ignore
\begin{abstract}

We report spectroscopic redshift measurements for three bright submillimeter galaxies (SMGs) near the GOODS-N field, each with SCUBA-2 850 \micron{} fluxes $> 10$ mJy, using the Northern Extended Millimeter Array (NOEMA). Our molecular linescan observations of these sources, which occupy an $\sim 7$~arcmin$^2$ area outside of the \textit{HST} coverage of the field, reveal that two lie at $z \sim 3.14$. In the remaining object, we detect line emission consistent with CO(7--6), [C \textsc{i}], and H$_2$O at $z = 4.42$. The far-infrared spectral energy distributions of these galaxies, constrained by SCUBA-2, NOEMA, and {\em Herschel}/SPIRE, indicate instantaneous star formation rates $\sim4000 ~{\rm M_{\odot}~yr^{-1}}$ in the $z=4.42$ galaxy and $\sim 2500~{\rm M_{\odot}~yr^{-1}}$ in the two $z\sim3.14$ galaxies. Based on the sources' CO line luminosities, we estimate $M_{{\rm gas}}\sim10^{11} M_{\odot}$ and find gas depletion timescales of $\tau_{{\rm depl}}\sim 50$ Myr, consistent with findings in other high-redshift SMGs. Finally, we show that the two $z\sim3.14$ sources, which alone occupy a volume $\sim10$ Mpc$^3$, very likely mark the location of a protocluster of bright SMGs and less dusty optical sources.

\end{abstract}
%TC:endignore



\section{Introduction} \label{sec:intro}

Submillimeter galaxies (SMGs) are home to some of the most extreme regions of star formation in the Universe. These highly dust-obscured sources, with far-infrared (FIR) luminosities $L_{\rm IR}$ in excess of $10^{12}~ L_{\odot}$, have star formation rates (SFRs) of hundreds to thousands of $\mathrm{M_{\odot}~ yr^{-1}}$ and typically lie at redshifts $z = 2 - 3$ \citep[e.g.,][]{Chapman05, Simpson14, Neri20}, though a significant tail in their redshift distribution has been found out to $z > 6$ \citep[e.g.,][]{Daddi09a, Daddi09b, Riechers20}. SMGs are major contributors to the SFR density of the early Universe, accounting for as much as half of all star formation at $z > 1$ \citep[e.g.,][]{Cowie17, Dud20}. The rapid buildup of stellar mass that results from their prodigious SFRs also suggests that SMGs are likely progenitors of compact quiescent galaxies at moderate redshifts and of massive ellipticals locally \citep[e.g.,][]{Simpson14, Toft14}. 

In addition to being an important phase in massive galaxy evolution, dusty starbursts may also trace the most massive dark matter halos in the early Universe \citep[e.g.,][]{Chen16, Dud20, Long20}. In the last ten years, a growing number of $z > 2$ overdensities and protoclusters of galaxies have been discovered through an excess of SMGs and luminous active galactic nuclei (AGNs), each containing several (sometimes $> 10$) such sources  \citep[e.g.,][]{Chapman09, Daddi09a,Daddi09b, Capak11, Walter12, Casey15, Miller18, Oteo18, GomezGuijarro19, Hill20, Long20, Riechers20, Zhou20}. 

However, the relatively short duration \citep[$\lesssim 100$ Myr, e.g.,][]{Carilli13} of a dusty starburst phase means that such structures may pose challenges to our understanding of galaxy evolution. Perhaps activation of the SMG phase is somehow correlated over the volume of a protocluster, though it is unclear how the canonical mechanisms for triggering a starburst in SMGs---gas-rich mergers \citep[e.g.,][]{Toft14} or the smooth accretion of gas from the surrounding medium \citep[e.g.,][]{Tadaki18}---could synchronize over large distances and short timescales. Alternatively, the gas depletion timescales $\tau_{\rm{depl}}$ in these environments may be significantly longer than previously thought, up to $\sim$1 Gyr in duration \citep[e.g.,][and references therein]{Casey16}. However, this conflicts with observations that suggest the most massive ellipticals in low-redshift clusters formed most of their stellar mass in short ($\lesssim$~1 Gyr) bursts at high redshift \citep{Thomas10}. Long $\tau_{\rm{depl}}$ would also imply impossibly massive end-product galaxies, if an SMG were to sustain its $\gtrsim 500~ M_{\odot}$ yr$^{-1}$ SFR for 1 Gyr. 

In any case, the existence of SMG-rich structures in the early Universe provides new and interesting constraints on the growth of individual massive galaxies and the assemblage of large-scale structures. Moreover, the diversity of observed properties of distant protoclusters \citep[e.g.,][and references therein]{Casey16} illustrates the need for a statistical sample of such structures if we are to make robust inferences about the growth of galaxy clusters and of massive galaxies across cosmic time.

In the first paper of their SUPER GOODS series, \citet{Cowie17} presented a deep 450~\micron{} and 850~\micron{} survey of the region around the GOODS-N field using SCUBA-2 on the 15~m James Clerk Maxwell Telescope (JCMT), along with interferometric followup of most of the more luminous SMGs with the Submillimeter Array (SMA). (Note that observations of the field have continued since that published work, and we use the latest images when quoting SCUBA-2 flux densities below.) 

Of the six SCUBA-2 sources in the field with 850~\micron{} fluxes greater than $\sim$10 mJy, four reside in a small ($\sim7$ arcmin$^2$) region to the northwest of the field center, just outside the \textit{HST}/ACS and WFC3 footprints of the GOODS \citep{goods04} and CANDELS \citep{Grogin11, Koek11} surveys; see the upper-right panel of Figure \ref{fig:sourcemap}. This includes one of the brightest SCUBA-2 850~\micron{} sources in the entire extended GOODS-N with an 850~\micron{} flux density of 18.7~mJy; cf. GN20 \citep{Pope05} at 16.3~mJy.

The three brightest SMGs in this grouping appear to be single sources at the resolution of the SMA observations, with the fourth only recently observed (M. Rosenthal et al., in preparation). More than being projected neighbors, their similar 20 cm flux densities, $K_s$ magnitudes, and photometric redshifts $z_{phot} \sim 3$ \citep{Yang14,Hsu19} suggest they may also lie at similar redshifts and may even belong to a single, massively star-forming structure. Two more moderately-bright SCUBA-2 sources, each with $S_{850} >$8~mJy and $z_{phot} \sim 3$, also lie in this region.

In this work, we present the first results of a spectroscopic campaign with the IRAM Northern Extended Millimeter Array (NOEMA) to detect CO line emission towards SMGs in this northwest offshoot of the GOODS-N. From these data, we determined spectroscopic redshifts of the three brightest sources and confirmed that two are at nearly identical redshifts. In Section \ref{sec:data}, we describe our NOEMA observations and data reduction, along with public multiwavelength imaging for our field. We present the redshifts and observed continuum and line properties of our sources in Section \ref{sec:results}. In Section \ref{sec:discussion}, we discuss the nature of these sources as well as their physical properties derived from our NOEMA data and from optical-through-mm spectral energy distribution (SED) fitting. We also discuss the evidence for these sources signposting a galaxy overdensity, concluding that they very likely belong to a protocluster that is relatively rich in SMGs. Finally, we give a brief summary in Section \ref{sec:summary}. Throughout this work we use a $\Lambda$CDM cosmology with $H_0=70.5$ km s$^{-1}$, $\Omega_{{\rm m}} = 0.27$, and $\Omega_\Lambda = 0.73$.

\section{Data} \label{sec:data}

\subsection{NOEMA Observations}

\begin{figure*}[th]
    \centering
    \includegraphics[width=0.85\linewidth]{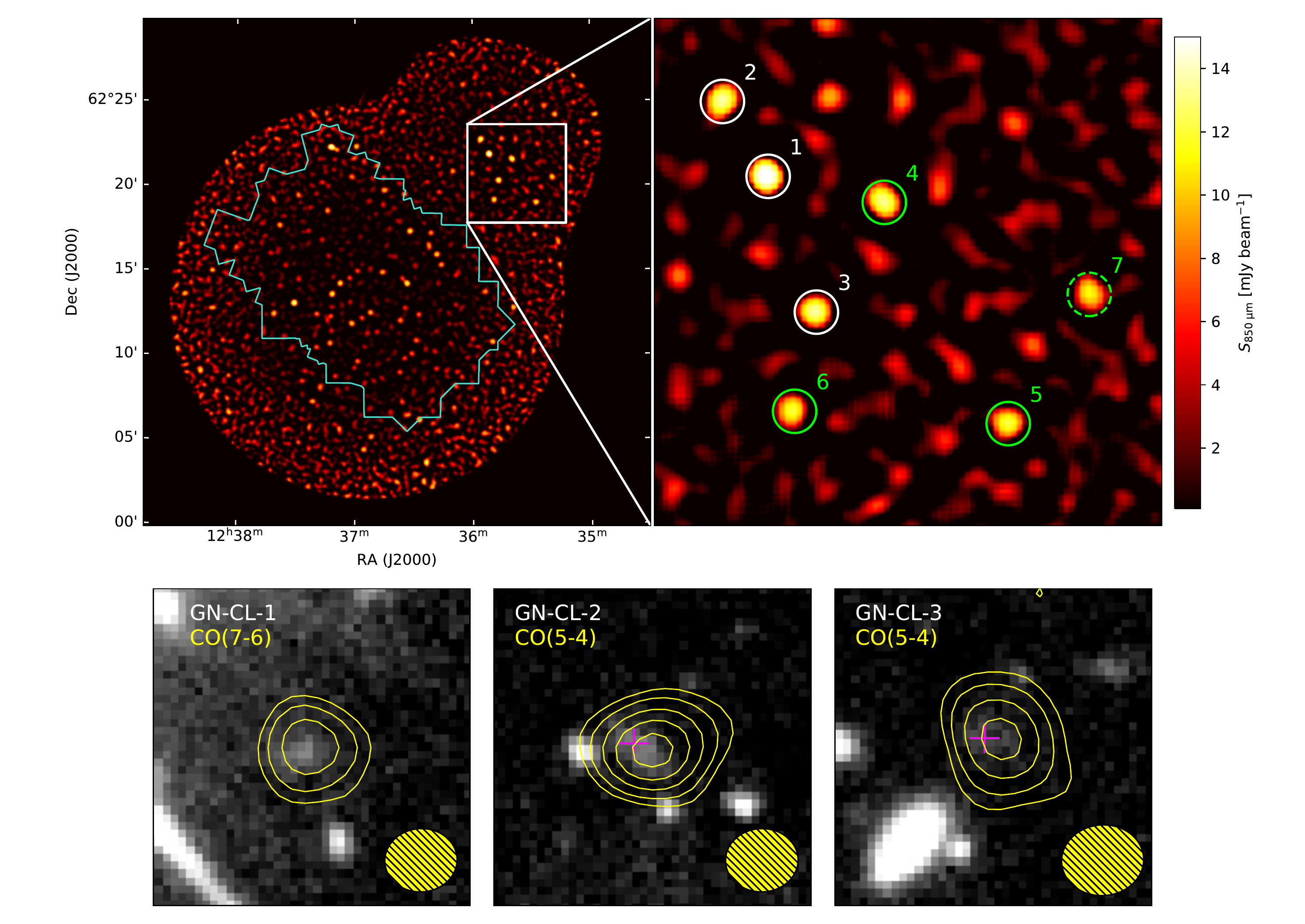}
    \caption{{\em Top left}: SCUBA-2 850 $\mu$m image of the extended GOODS-N. Displayed fluxes include a multiplicative factor of 1.1 to correct for the SCUBA-2 PSF. The blue contour shows the CANDELS {\em HST}/WFC3 F160W footprint. {\em Top right}: A zoom-in to the portion of the field that contains our target galaxies (marked by the white box in the top left panel). Our 3 target galaxies are circled in solid white. An additional 3 galaxies with $S_{850} >8$~mJy, which may also belong to the $z \sim 3.1$ protocluster, are circled in solid green. {\em Bottom}: CFHT/WIRCam $K_s$-band images of GN-CL-1, GN-CL-2, and GN-CL-3. Each panel is 12\arcsec~on a side. Pink crosshairs mark the position of each object's nearest $K$-band counterpart (except for GN-CL-1, which has no catalogued counterpart in \citet{Hsu19} due to its proximity to a bright star). Contours show the 3, 7, 12, 18, and 25$\sigma$ levels of the CO(7--6) emission line in GN-CL-1 and CO(5--4) in GN-CL-2 and GN-CL-3. In the bottom-right corner of each panel, we show the beam profiles for the CO maps as yellow ellipses.}
    \label{fig:sourcemap}
\end{figure*}

We targeted the three brightest SCUBA-2 sources in the northwest offshoot of the GOODS-N using the NOEMA interferometer in the compact D configuration (project ID W19DG; PI: Jones). We used two spectral tunings each in the \textit{PolyFix} 2~mm and 3~mm bands, which cover the frequency ranges 78.384--109.116 GHz and 131.384--162.116 GHz. GN-CL-1 and 2 were observed only in the 2 mm band, as 3 mm observations towards these sources had already been carried out in 2019 August; to our knowledge, these data have not yet been published. GN-CL-3 was observed with all four spectral setups. Observations were carried out in track-sharing mode in good weather conditions over the course of 2020 April 10--16, with average atmospheric phase stability of $\sim10$--30\% degrees rms and typical precipitable water vapor levels of 1--4 mm. Tracks executed on 2020 April 10 used nine antennas, while all others used ten. In all observations, the quasar 1030+611 was used as the phase and amplitude calibrator. Observations carried out on 2020 April 10 used 0851+202 as the flux calibrator, while all others used LkH$\alpha$101. Calibration and imaging of the \textit{uv} data were carried out in \gildas. We estimate that the absolute flux calibration is accurate at the $\sim$15\% level. Images were produced using natural weighting, with typical synthesized beam sizes of $7'' \times 4''$ ($3'' \times 2\farcs5$) at 3 (2) mm (see bottom panels in Figure \ref{fig:sourcemap}).

Identification of lines in each tuning and sideband, as well as separation of line- and continuum-only information, were carried out using an iterative process. First, cleaned spectral cubes were binned to $\sim$75 km s$^{-1}$ channel widths to better identify potential emission and absorption features. Strong emission features were identified by eye and then masked with the \textsc{uv\_filter} task in \gildas/\textsc{mapping} using a $\sim$800--1000~km~s$^{-1}$ wide window around the frequency of the line peak. This somewhat aggressive method of line-masking ensures that our continuum measurements remain uncontaminated by strong emission features at the cost of slightly underestimating the continuum flux density. The remaining channels were then collapsed to form our continuum-only images. RMS noise values in a given window were essentially uniform for all sources observed in that window, ranging from 16--19.8~$\mu$Jy in the 3~mm band and 23.2--30.1~$\mu$Jy in the 2~mm band. Continuum-subtracted spectral cubes were created with the \textsc{uv\_base} task in \gildas/\textsc{mapping} using the same windows as described previously to mask out strong lines.

\subsection{Multiwavelength Data}

\begin{table*}[th]
    \centering
    \caption{{\sc NOEMA Line and Continuum Measurements}}
    \begin{tabular}{cccccccccccc}
    \hline
    \hline
    Name & R.A.$^a$ & Dec.$^a$ & ${S_{\rm 850\mu m}}^b$ & ${S_{158.2 {\rm GHz}}}^c$ & Line & $\nu_{{\rm obs}}$ & $z$ & $\nu_0$ & $\Delta v_{\rm FWHM}$ & $S_{{\rm peak}}$ \\
    & & & [mJy] & [mJy] &  & [GHz] &  & [GHz] & [km s$^{-1}$] & [mJy beam$^{-1}$] \\ 
    \hline
    GN-CL-1 & 188.96404 & 62.36311 & 18.7$\pm$0.5 & 2.40$\pm$0.03 & CO(7--6) & 148.76 & 4.422 & 806.65 & 513$\pm$49 & 3.1$\pm$0.3 \\
    & & & & & [C~{\sc i}]($^3$P$_2$--$^3$P$_1$) & 149.24 & 4.423 & 809.34 & 362$\pm$63 & 2.0$\pm$0.3 \\
    & & & & & H$_2$O($2_{11}$--$2_{02}$) & 138.72 & 4.421 & 752.03 & 430$\pm$85 & 1.2$\pm$0.2 \\
    % SMA123551622147
    \hline
    GN-CL-2 & 188.98283 & 62.37750 & 11.2$\pm$0.5 & 0.87$\pm$0.03 & CO(5--4) & 138.92 & 3.148 & 576.27 & 369$\pm$23 & 4.3$\pm$0.2 \\
    % SMA123555622239
    \hline
    GN-CL-3 & 188.94433 & 62.33703 & 11.5$\pm$0.6 & 0.90$\pm$0.03 & CO(5--4) & 139.46 & 3.132 & 576.27 & 500$\pm$41 & 3.0$\pm$0.2 \\
    & & & & & CO(3--2) & 83.71 & 3.131 & 345.80 & 459$\pm$40 & 2.0$\pm$0.2 \\
    % SMA123546622013
    \hline
    \end{tabular}
\label{tab:sources}
\tablecomments{$^a$Source positions are from the SMA observations of \citet{Cowie17} (columns 8 and 9 of their Table 5).\\
$^b$The 850~$\mu$m fluxes are from the latest SCUBA-2 images.\\
$^c$This sideband is devoid of obvious spectral features in all of our sources, providing a clean continuum measurement. The flux uncertainties are from the 2D Gaussian fits to the continuum images and do not include systematic uncertainties, such as from absolute flux calibration, which may be $\sim$15\%.}
\end{table*}

% Optical and NIR
Because \textit{HST} data are not available for our sources, we instead used the compilation of deep ultraviolet (UV), optical, and near-infrared (NIR) photometry of the extended GOODS-N from \citet{Hsu19} and references therein to constrain the stellar properties of our sources. Specifically, for our SED fits (see Section~\ref{subsec:seds}), we used Subaru/Suprime-Cam \textit{BVRIz} data from \citet{Capak04};
CFHT/WIRCam \textit{JHK$_s$} data from \citet{Wang10} and \citet{Hsu19}; and \textit{Spitzer}/IRAC 3.6~\micron{} and 4.5~\micron{} data from \citet{Ashby13}. Two of our three sources (GN-CL-2 and 3) have counterparts in the \citet{Hsu19} catalog within 1\arcsec~of the SMA 870~\micron{} centroid, and we used their photometry directly for these sources. GN-CL-1 is very near a bright ($H = 7.8$) star and thus no nearby optical-NIR counterpart is listed in the \citet{Hsu19} catalog. Instead, we performed our own aperture photometry on the \textit{JHK$_s$} data from \citet{Wang10} and \citet{Hsu19} at the SMA position of GN-CL-1. In each band, we measured fluxes in a 2\arcsec\ diameter aperture and a local median ``background" (which largely comes from the saturated foreground star) in a 2.4\arcsec--6\arcsec\ diameter annulus, both centered on the SMA position of GN-CL-1. We use the median background values to correct our fluxes for spillover light from the star.

% FIR to radio
At long wavelengths, we use data from the GOODS-{\em Herschel} program of \citet{Elbaz11} to measure SPIRE 250~$\mu$m, 350~$\mu$m, and 500~$\mu$m fluxes for our sources. Finally, we use the Very Large Array (VLA) 20 cm observations of \citet{Morrison10} to search for radio counterparts, though the radio fluxes are not included in our SED fits.

\section{Results}
\label{sec:results}

\begin{figure*}
    \centering
    \includegraphics[width=0.9\linewidth]{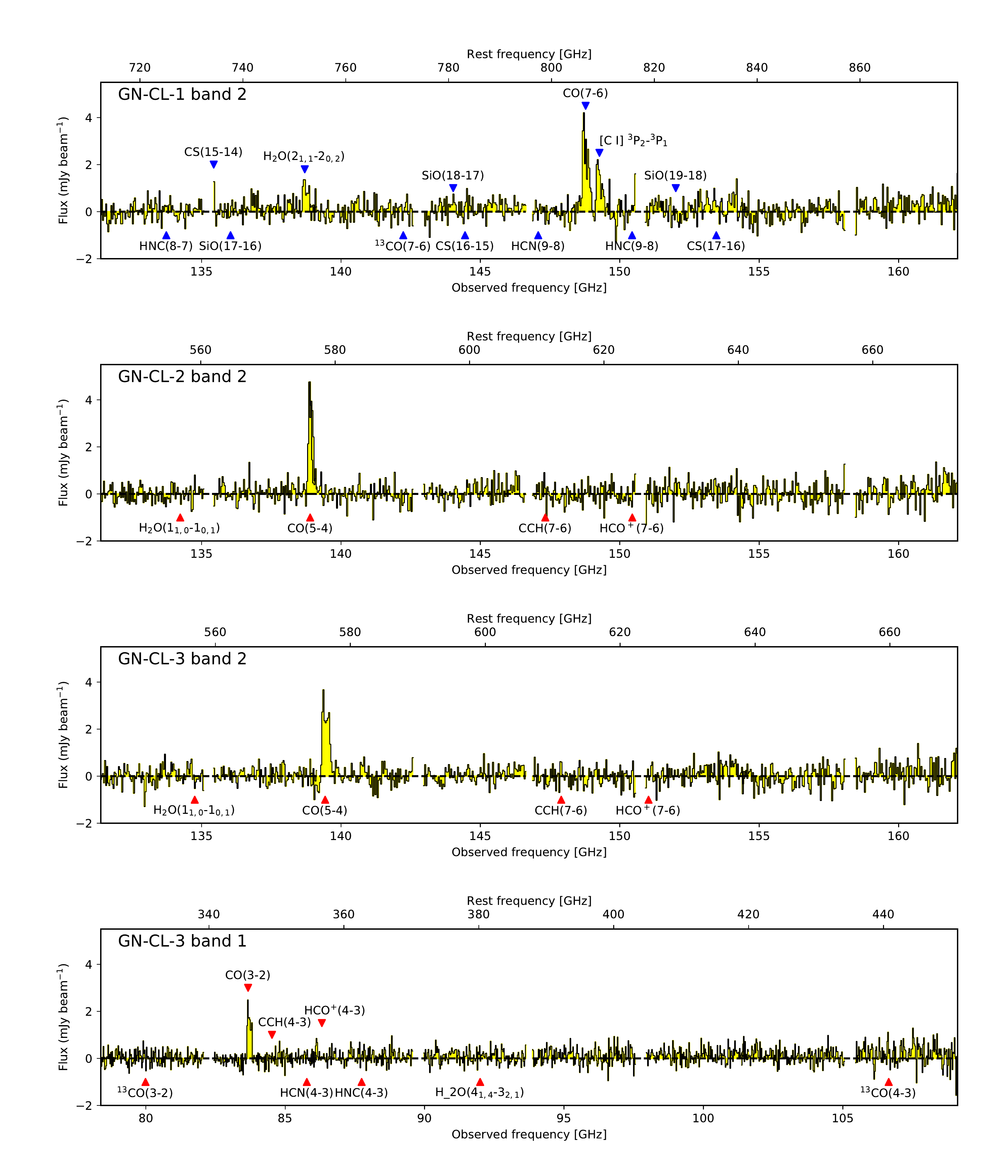}
    \caption{Full millimeter spectra of our three bright SCUBA-2 sources. The source name and band are labeled in the top left corner of each panel. Continuum-subtracted spectra were extracted from the spaxel with the brightest continuum emission. The triangles denote the frequencies of common molecular and atomic emission and absorption lines in the spectral ranges of our observations. Given the rms noise in these spectra, we can only securely identify the CO, H$_2$O, and [C {\sc i}] emission lines listed in Table \ref{tab:sources}. However, for completeness, we also label several weaker transitions that are known to exist in these ranges. Red triangles show commonly observed transitions in SMGs from \citet{Spilker14}. Because that work does not extend to the highest rest-frame frequencies at which we observed GN-CL-1, we mark transitions from the NRAO's Splatalogue database as blue triangles on its spectrum.}
    \label{fig:full_spectra}
\end{figure*}

For GN-CL-1 and GN-CL-2, we extract four continuum flux densities in windows centered at 135.3, 142.8, 150.7, and 158.2~GHz. For GN-CL-3, we extract four at the above frequencies and another four in windows centered at 82.3, 89.8, 97.7, and 105.2 GHz. All of our sources are securely detected in continuum (at the $\gtrsim4\sigma$ level) in all sidebands of all tunings in which they were observed, though for brevity, we report only the 158.2~GHz flux densities in Table \ref{tab:sources}, as this sideband is devoid of obvious line emission or absorption in all of our sources and thus provides a clean continuum measurement. As may be expected from their relative 850~\micron{} fluxes, the 158.2~GHz flux densities of GN-CL-2 and 3 are similar at around 0.9~mJy, while that of GN-CL-1 is brighter; we give exact values in Table \ref{tab:sources}.

\begin{figure}
    \centering
    \includegraphics[width=0.85\linewidth]{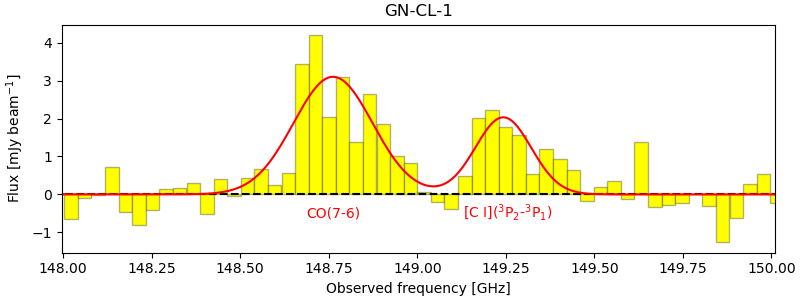}
    \includegraphics[width=0.85\linewidth]{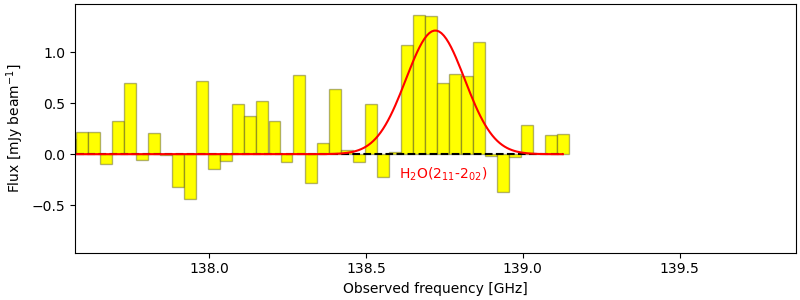}
    \includegraphics[width=0.85\linewidth]{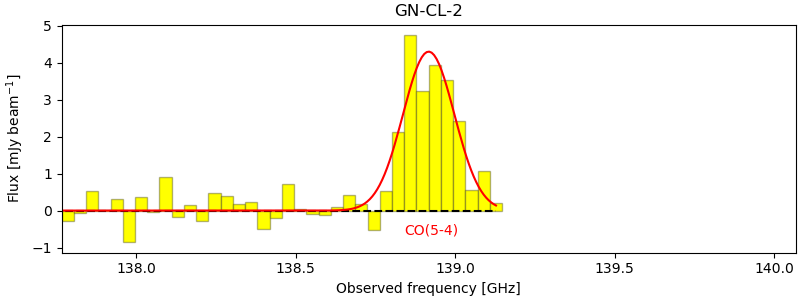}
    \includegraphics[width=0.85\linewidth]{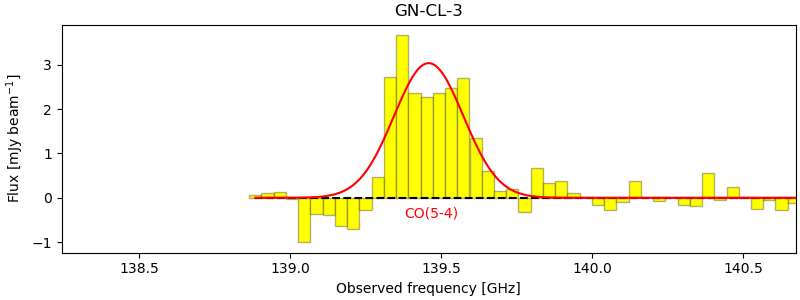}
    \includegraphics[width=0.85\linewidth]{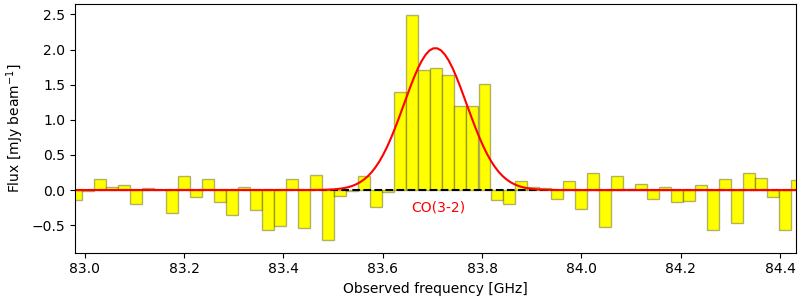}
    \caption{Continuum-subtracted CO or H$_2$O emission from our three bright SCUBA-2 sources, zoomed in from the full spectra show in Figure \ref{fig:full_spectra}. Spectra are extracted from the spaxel with the brightest continuum emission. Red lines show Gaussian fits to the data, with the transition labeled in red below the line. The fitted peak flux densities, FHWM velocities, and peak frequencies are listed in Table \ref{tab:sources}.}
    \label{fig:spectra}
\end{figure}

We show the full NOEMA spectra of each of our sources in Figure \ref{fig:full_spectra}, and mark common molecular and atomic transitions in these bands. We do not detect most of the weaker emission features, but GN-CL-1, 2, and 3 all have at least one millimeter emission line detected at $>$5$\sigma$. We show the spectra from the spaxel with the brightest line emission in Figure \ref{fig:spectra}. Below we discuss the continuum flux densities, line properties, and redshifts of the sources individually.

\underline{\em GN-CL-1}: This source is extremely well-detected in continuum, with flux densities rising from 1.3 $\pm$ 0.03 mJy (SNR$\sim$43) in the 135.3~GHz window to 2.4$\pm$0.03 mJy (SNR$\sim$80) in the 158.2~GHz window. We detect two strong emission features towards this source at 148.762 and 149.243~GHz and a weaker emission feature at 138.721~GHz, consistent with the frequency ratios of CO(7--6), [C~{\sc i}]($^3$P$_2$--$^3$P$_1$), and H$_2$O($2_{11}$--$2_{02}$) at $z\approx4.42$. We fit the CO(7--6) and [C~{\sc i}]($^3$P$_2$--$^3$P$_1$) lines simultaneously with two Gaussians, without fixing their line ratios, widths, or frequencies relative to one another, and we fit a single Gaussian to the H$_2$O feature. The three lines have peak flux densities of $3.1\pm0.3$, $2.0\pm0.3$, and $1.2\pm0.2$ mJy beam$^{-1}$ for CO(7--6), [C~{\sc i}]($^3$P$_2$--$^3$P$_1$), and H$_2$O($2_{11}$--$2_{02}$), respectively, with line widths ranging from $\sim$360 to 510~km~s$^{-1}$. As we discuss below, this redshift identification shows that GN-CL-1 is not physically associated with either of the remaining two sources.

\begin{figure}[bht]
    \centering
    \includegraphics[width=0.9\linewidth, trim={0.35cm 0.2cm 1.4cm 0.7cm},clip]{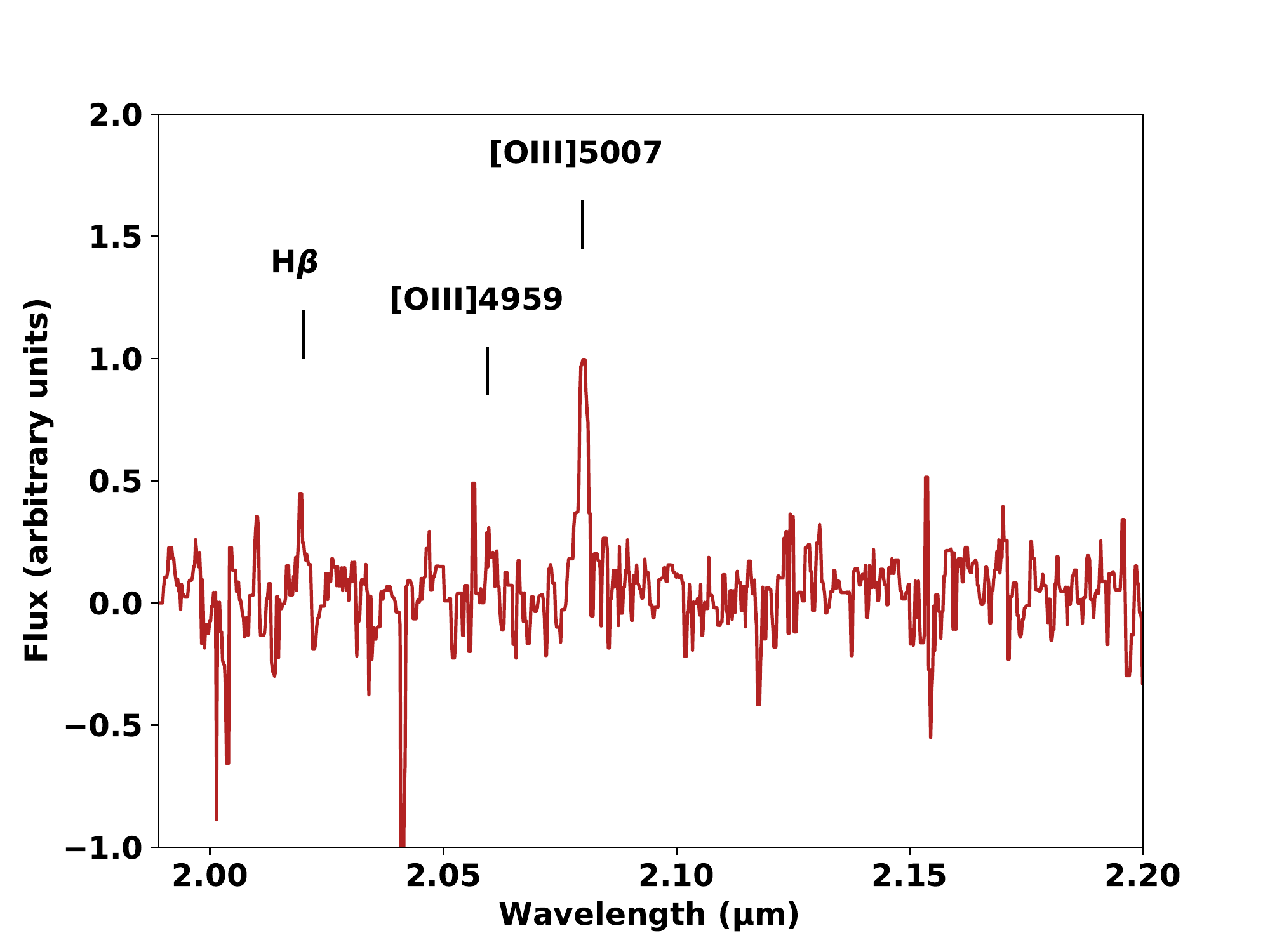}
    \caption{Portion of the Keck/MOSFIRE $K$-band spectrum of optical source 77630 in the catalog of \citet{Hsu19}, the most likely optical-NIR counterpart to GN-CL-2. The data have been median filtered using a window that is 5 wavelength bins wide. We mark the observed positions of the redshifted [O{\sc iii}] doublet and H$\beta$ at $z = 3.15$.}
    \label{fig:mosfire_cl2}
\end{figure}

\underline{\em GN-CL-2}: The second brightest SCUBA-2 source in our sample has NOEMA continuum flux densities ranging from 0.41$\pm$0.02 mJy in the 135.3~GHz window to 0.87$\pm$0.03 mJy in the 158.2~GHz window. We detect a single strong emission line towards GN-CL-2 at 138.917~GHz, with a peak flux density of $4.3\pm0.2$ mJy beam$^{-1}$ and a FWHM of $\approx$~370~km~s$^{-1}$.
%The remarkable similarity of its FIR SED with that of GN-CL-3 implies a secure redshift of $2 < z < 5$.
For this redshift range, based on the most common millimeter transitions in high-redshift SMGs \citep[e.g.,][]{Spilker14}, a line of this strength is most likely to be CO with $J_{\rm up} =$ 4, 5, 6, or 7 at $z =$ 2.32, 3.15, 3.98, or 4.81, respectively, [C {\sc i}](1--0) at $z=2.54$, or H$_2$O(2$_{11}$--2$_{02}$) at $z = 4.41$.

Our near-continuous coverage from 131.4--162.1~GHz allows us to rule out most of these redshift identifications by the non-detection of other strong lines at the expected frequencies. If the feature at 138.917 GHz were CO(4--3) at $z = 2.32$, for example, we would expect to detect [C{\sc i}](1--0) at $\nu_{\rm obs} \approx 148.2$ GHz, but no significant line emission is seen there. Redshifts of $z = 3.98$, 4.41, and 4.81 are similarly ruled out by the absence of strong ($\gtrsim1$ mJy beam$^{-1}$) H$_2$O($2_{11}-2_{02}$), CO(7--6), and [C\textsc{i}]($^3$P$_2$--$^3$P$_1$) emission at their respective expected frequencies.

This leaves CO(5--4) at $z = 3.15$ and [C{\sc i}](1--0) at $z = 2.54$ as the only viable redshift identifications from the NOEMA data alone. Follow-up Keck/MOSFIRE $K$-band spectroscopy from M. Rosenthal et al., in preparation, of this source's nearest optical-NIR counterpart (source 77630 in \citealt{Hsu19}, with $K = 22.7$ and a separation of 0\farcs{709} from the SMA position of GN-CL-2) finds a redshift of $z$ = 3.15 based on the [O{\sc iii}] doublet and H$\beta$ (see Figure \ref{fig:mosfire_cl2}). This confirms the higher redshift identification.
%of this source's nearest optical-NIR counterpart (source 77630 in \citet{Hsu19}, with $K = 22.7$ and a separation of 0\farcs{709} from the SMA position of GN-CL-2) reveals a three lines at approximately 2.02, 2.06 and 2.08 $\mu$m; see Figure \ref{fig:mosfire_cl2}. The wavelength separation and relative line strengths are consistent with being the [O{\sc iii}] doublet and H$\beta$ at $z \approx 3.15$, while no lines are expected at these wavelengths from a galaxy at $z = 2.54$. 

\underline{\em GN-CL-3}: GN-CL-3 was the only source we observed in both the 2~mm and 3~mm bands. The nearest optical-NIR counterpart to this SMG (separation of 0\farcs{473} from its SMA position) is source 85384 in \citet{Hsu19}. Its millimeter continuum flux densities rise smoothly from 82.5$\pm$16~$\mu$Jy in the 82.3~GHz window to 0.9$\pm$0.03~mJy in the 158.2~GHz window. We detect strong emission features at 83.706~GHz and 139.458~GHz, consistent with the frequency ratio of CO(3--2) and CO(5--4) at $z \approx 3.13$. Separate Gaussian fits to each of these lines yield peak flux densities of $2.0\pm0.2$ and $3.0\pm0.2$ mJy beam$^{-1}$, respectively, with FWHM $\approx$~480~km~s$^{-1}$. GN-CL-2 and GN-CL-3 are therefore confirmed to lie a mere $\Delta z\sim0.02$ apart in redshift, with a projected separation of 1.2~Mpc and a 3D separation of 2.7 proper (14.6 comoving)~Mpc, if the difference in recession velocities is due only to the Hubble flow. A sphere of diameter equal to the proper distance between these two galaxies would have a volume of, at most, $\sim$10~Mpc$^3$. However, as we discuss in the next section, these galaxies very likely belong to a larger overdensity of rare, submillimeter-bright galaxies. If the difference in their redshifts is due to peculiar velocities within a common structure, then the separations quoted above would be overestimates.

Finally, we note that GN-CL-3 appears to have a projected companion about 5\arcsec~to the southeast. This neighboring source is extremely faint in continuum, with a $< 2\sigma$ detection in the $\sim2$~mm data and completely invisible in the lower-resolution 3~mm data. However, its presence is revealed by a single strong emission feature (peak flux density $\sim 2.5$~mJy) at 149.3~GHz. No line emission from an SMG at $z=3.13$ is expected at this frequency, which suggests that the companion is not associated with GN-CL-3. The line emission is spatially coincident with an optically-bright $z=0.543$ radio source \citep{Barger14}, which suggests this line may be CO(2--1).

\begin{table}[b!]
    \centering
    \caption{{\sc Derived Properties of Candidate Protocluster Members}}
    \begin{tabular}{cccc}
        \hline
        \hline
         & GN-CL-1 & GN-CL-2 & GN-CL-3 \\
        \hline
        SFR$^a$ [${\rm M_\odot~ yr}^{-1}$] & $3900\pm580$ & $2530\pm130$ & $2990\pm770$ \\
        ${\rm SFR_{100}}^b$ [${\rm M_\odot~ yr}^{-1}$] & $1070\pm300$ & $630\pm30$ & $390\pm140$ \\
        $M_\star$ [$10^{10}~{\rm M_\sun}$] & $10.7\pm3.1$ & $5.8\pm0.3$ & $4.5\pm1.2$ \\
        sSFR [${\rm Gyr^{-1}}$] & $37\pm12$ & $44\pm3$ & $67\pm25$ \\
        $M_{\rm d}$ [$10^9~{\rm M_\sun}$] & $4.5\pm0.4$ & $2.2\pm0.2$ & $2.0\pm0.1$ \\
        $T_{\rm d,char}$ [K] & $33.4\pm0.8$ & $34.9\pm0.5$ & $34.8\pm0.6$ \\
        $M_{\rm gas}$ [$10^{11}~{\rm M_\sun}$] & $2.0\pm0.5$ & $1.3\pm0.2$ & $1.2\pm0.2^c$ \\
        ${\tau_{\rm depl}}$ [Myr] & $52\pm15$ & $51\pm9$ & $41\pm13^c$ \\
        $M_{\rm gas}/M_{\rm d}$ & $45\pm12$ & $59\pm11$ & $61\pm12^c$ \\
        \hline
    \end{tabular}
    \tablecomments{\\$^a$Instantaneous SFR from {\sc cigale}. This row is used to calculate sSFR and $\tau_{\rm depl}$. \\ 
    $^b$100 Myr-averaged SFR from {\sc cigale}. \\
    $^c$GN-CL-3 values use $M_{\rm gas}$ from CO(5--4), but these are consistent within error with values from CO(3--2) (see Section \ref{subsec:gas}).}
    \label{tab:derived_properties}
\end{table}

\section{Discussion} \label{sec:discussion}

\subsection{Star Formation Rates, Masses, and Dust Temperatures} \label{subsec:seds}

The rich photometric data from the rest-frame optical to the radio allow us to constrain the physical properties of these galaxies with SED fitting. We use the Code Investigating GALaxy Emissions \citep[{\sc cigale},][]{Noll09}, which calculates SEDs using an energy balance principle, where the energy absorbed by dust in the UV to NIR equals that re-radiated in the MIR to FIR. We use the updated {\sc python} version of the code, which has been shown to produce comparable results for high-redshift starbursts to other SED fitting codes \citep{Boquien19}, to constrain the SFRs, stellar masses, dust masses, and dust temperatures of the three SMGs.

We used \citet{Bruzual03} stellar population libraries with a \citet{Chabrier03} initial mass function (IMF) for the full range of available metallicities. Stellar spectra are attenuated using a modified \citet{Charlot00} dust law with fixed power law indices $\delta_{\rm ISM} = -0.7$ and $\delta_{\rm BC}=-1.3$ and a separation age between old and young stellar populations of 10~Myr. We use the dust emission models from \citet{Draine14}, with an input minimum radiation field of $1.0 \leq U_{\rm min} \leq 50.0$\footnote{$U$ has units of 1 Habing $= 1.6\times10^{-3}~{\rm erg~cm^{-2}~s^{-1}}$} and an input mass fraction of $0.005 \leq \gamma \leq 0.05$ irradiated by $U > U_{\rm min}$. We fix the radiation field power law slope $\alpha=2.0$, and we fit for polycyclic aromatic hydrocarbon (PAH) mass fractions $0.47\% \leq q_{\rm PAH} \leq 3.90\%$. The mean intensity $\left<U\right>$ is used to derive a characteristic dust temperature, $T_{\rm d,char} = 18~{\rm K}\times \left<U\right>^{1/6}$ \citep{Draine14}, which we report in Table \ref{tab:derived_properties}.

The properties estimated by {\sc cigale}, especially the SFR, are strongly dependent on the input star formation history (SFH). Given these galaxies have FIR fluxes indicative of ongoing starbursts, we model the SFH as having formed 50-99\% of stars by mass in a short, ongoing, flat burst of age $\leq 100$ Myr, and the remainder of stars in an exponentially declining SFH of age 0.25-1.5~Gyr, with $e$-folding time $250~{\rm Myr} \leq \tau_{\rm main} \leq 6~{\rm Gyr}$. {\sc cigale} returns a maximum likelihood instantaneous SFR for each galaxy, as well as SFRs averaged over the preivous 10 and 100~Myr. Different methods of SFR estimation in the literature report either averaged or instantaneous SFRs, so we report both the instantaneous and 100 Myr-averaged SFR for each galaxy in Table \ref{tab:derived_properties}. We use the instantaneous SFRs to compute gas depletion times (see Section \ref{subsec:gas}). The 100~Myr-averaged SFR is appropriate for comparisons with SFRs output from codes such as {\sc magphys} \citep{daCunha15} and works that use it \citep[e.g.][]{Dud20}.

\begin{figure}
    \centering
    \includegraphics[width=0.9\linewidth,trim={0.5cm 0 1.1cm 0},clip]{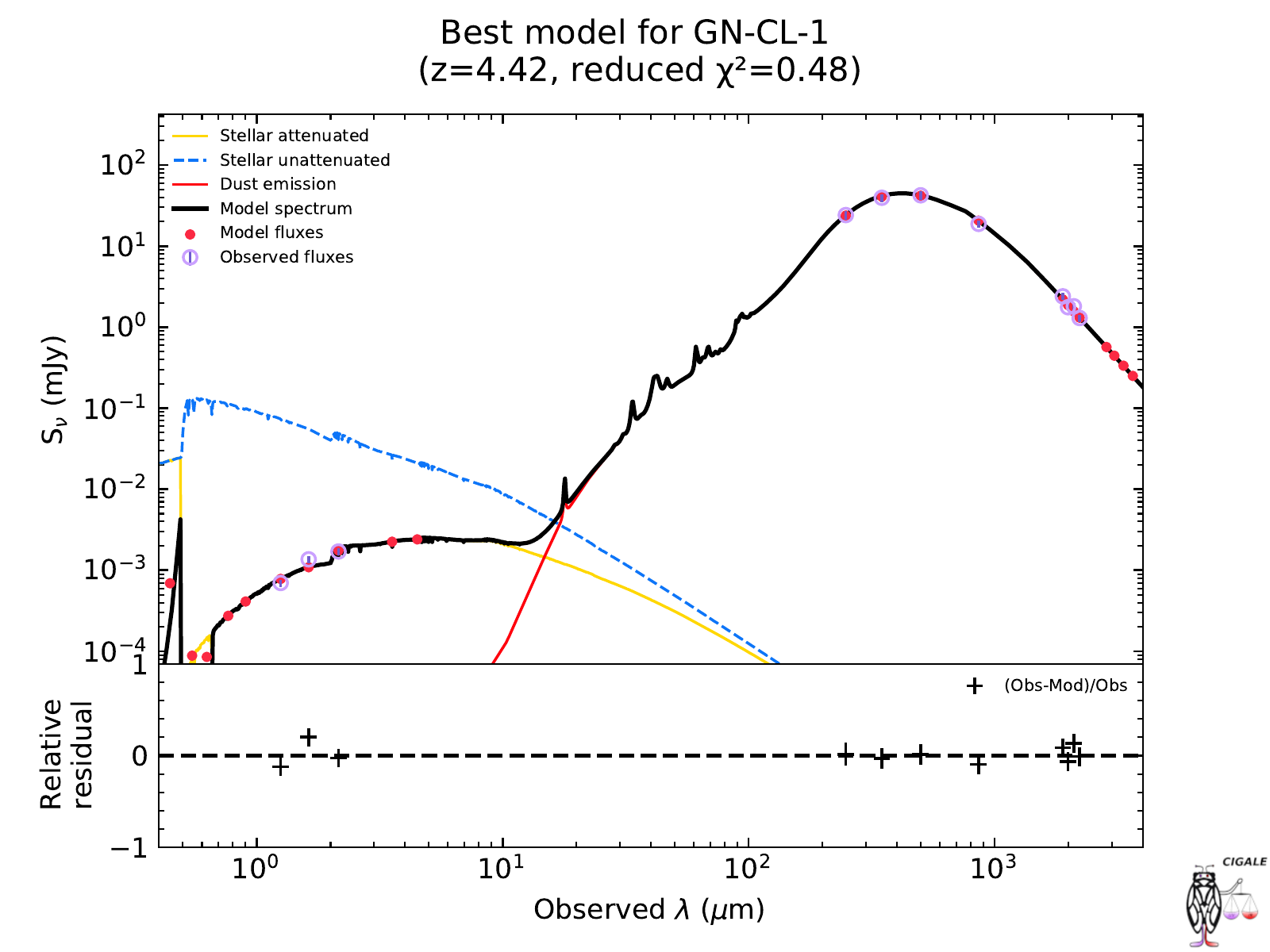}
    \includegraphics[width=0.9\linewidth,trim={0.5cm 0 1.1cm 0},clip]{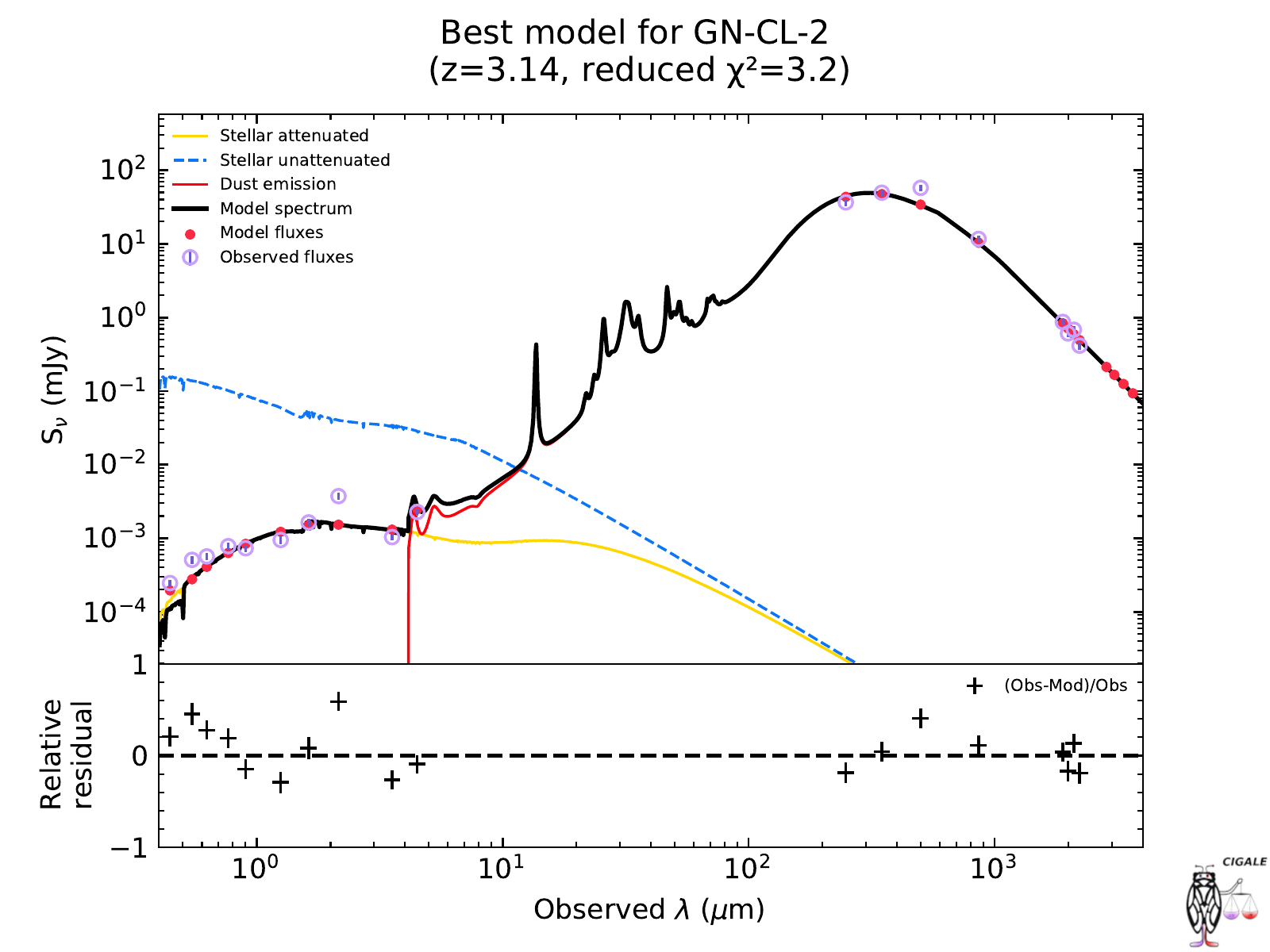}
    \includegraphics[width=0.9\linewidth,trim={0.5cm 0 1.1cm 0},clip]{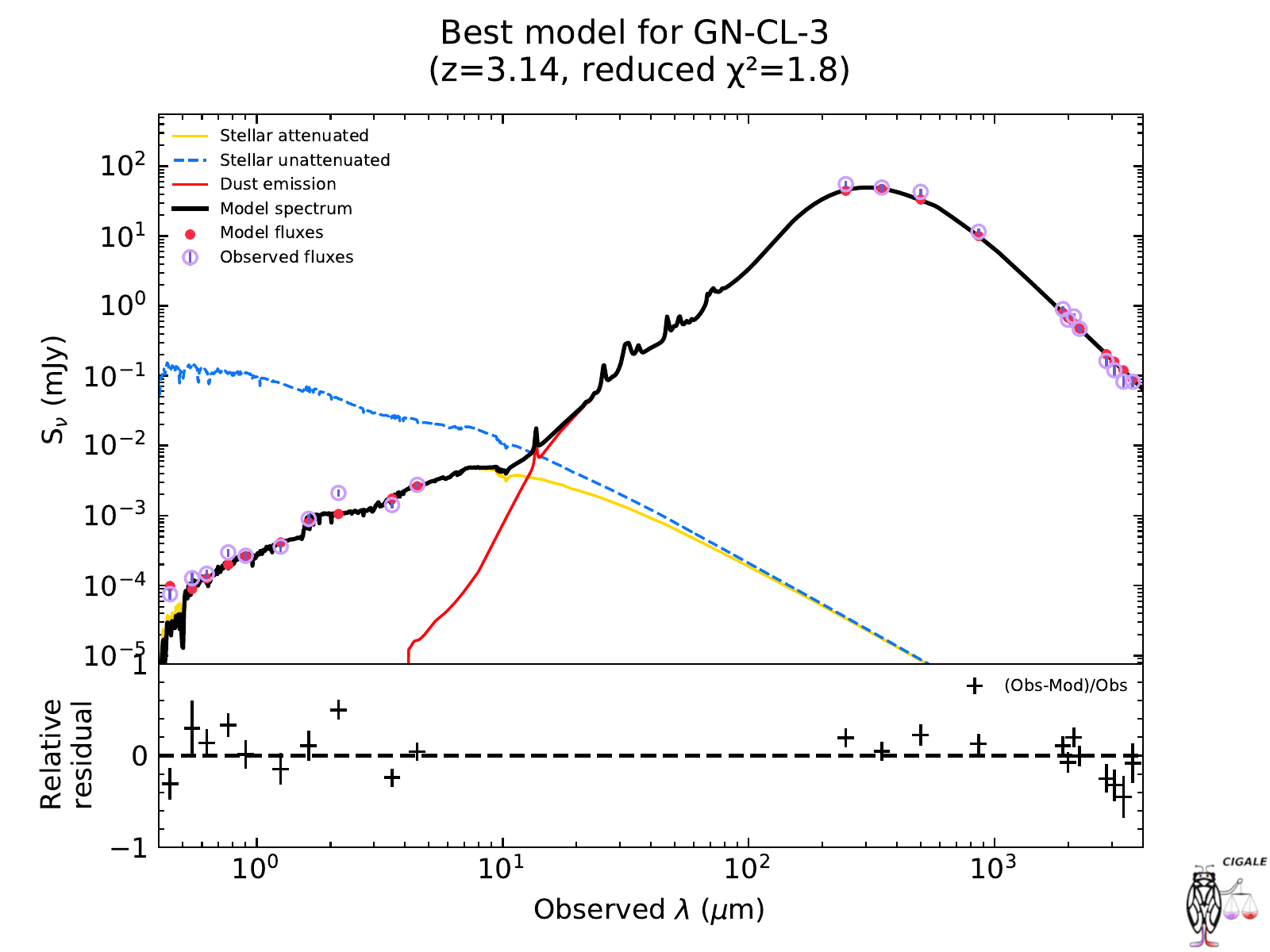}
    \caption{Best-fit SEDs from \textsc{cigale} for each SMG. The black curve shows the total SED, while the red and yellow curves show the relative contributions to the SED of the dust and the attenuated stellar emission, respectively. The purple and red dots show the measured and fit fluxes, respectively.}
    \label{fig:SEDs}
\end{figure}

We ran {\sc cigale} on FIR to millimeter photometry from {\em Herschel}/SPIRE, SCUBA-2, and NOEMA, along with observed-frame optical to NIR photometry from \citet{Hsu19} (for GN-CL-2 and 3) or our own {\em JHKs} aperture photometry (for GN-CL-1).
% We excluded the {\em U}-band photometry from the KPNO Mayall 4m, as none of our sources had $>3\sigma$ {\em U} detections.
%consistent with this wavelength corresponding to rest-frame wavelengths blueward of the Lyman-break.
In addition to the cataloged measurement uncertainties, we included a 10\% systematic uncertainty on the absolute photometry/flux calibration in our input flux errors. We fixed the redshifts at the spectroscopic redshift of each source; that is, $z_{spec}=4.42$ for GN-CL-1 and $z_{spec}=3.14$ for GN-CL-2 and GN-CL-3. We show the best-fit SEDs for the three SMGs in Figure \ref{fig:SEDs}.

% The stellar mass, SFR, dust mass, and characteristic dust temperature from SED fitting are given in Table \ref{tab:derived_properties}.
Based on its best-fitting SED, GN-CL-1 has an 8--1000~$\mu$m luminosity $L_{\rm IR} \approx 1.1\times10^{13}~{\rm L_\sun}$, making it a high-redshift hyper-luminous infrared galaxy (HyLIRG). Its instantaneous ${\rm SFR} = 3900 \pm 580~{\rm M_\sun~yr^{-1}}$ is one of the largest in the field. With $\sim10^{11}~{\rm M_\sun}$ of stars already in place by $z=4.42$, this source is the likely progenitor of a massive elliptical at $z\approx 0$. GN-CL-2 and GN-CL-3 are also massive star-forming galaxies, with 
8--1000~$\mu$m luminosity $L_{\rm IR} \approx 5\times10^{12}~{\rm L_\sun}$, comparable to nearby ultraluminous infrared galaxies, and instantaneous ${\rm SFR}\approx2500~{\rm M_\sun~yr^{-1}}$. The similarities between GN-CL-2 and GN-CL-3, as well as the difference in their values compared with GN-CL-1, are consistent with rough expectations from the 850~$\mu$m flux densities. %All three galaxies have ${\rm sSFR \gtrsim 40~Gyr^{-1}}$, which are significantly higher than expected for galaxies of their stellar masses at $z = 3 - 5$ \citep[e.g.][]{Tasca15}.

\subsection{Gas Mass and Depletion Timescale} \label{subsec:gas}

Molecular hydrogen's low emissivity makes it hard to detect even in nearby galaxies, so the luminosity of the CO(1--0) emission line is often used as a proxy for the mass of molecular hydrogen, $M_{{\rm H_2}}$. We follow the methodology of \citet{Bothwell13} to convert our higher $J$ CO transitions to CO(1--0), and subsequently into a molecular gas mass for each galaxy, $M_{{\rm gas}}$. 

We compute the line luminosity $L'_{\rm CO}$ using the standard relation from \citet{Solomon05}:
\begin{equation}
    L'_{\rm CO} = 3.25\times10^7 \times S_{\rm CO} \Delta V \nu_{\rm obs}^{-2} D_{\rm L}^2 (1+z)^{-3},
\end{equation}
where $L'_{\rm CO}$ is the line luminosity with units ${\rm K~km~s^{-1}~pc^2}$, $S_{\rm CO}\Delta V$ is the integrated line luminosity in Jy km s$^{-1}$, $\nu_{\rm obs}$ is the observed line frequency in GHz, and $D_{\rm L}$ is the luminosity distance in Mpc. We take $S\Delta V$ to be the area underneath the Gaussian fits to each detected CO line and convert our measured luminosities from higher $J$ lines to CO(1--0) using Table 4 of \citet{Bothwell13}.
%$D_{\rm L}$ is derived from the measured $z$, and errors on $z$ are $\sim$10$^{-5}$, so 
Errors on $L'_{\rm CO}$ are dominated by errors on $S\Delta V$ and on the $J$-conversion factors $r_{76/10}$, $r_{54/10}$, and $r_{32/10}$. We assume an $L'_{\rm CO(1-0)}$ to $M_{{\rm H_2}}$ conversion factor $\alpha = 1~{\rm M_\sun ( K~km~s^{-1}~pc^2})^{-1}$. Finally, we multiply a correction of $M_{\rm gas} = 1.36 M_{\rm H_2}$ to account for the addition of helium.

For a given SFR and $M_{\rm gas}$, the depletion timescale is approximately given by 
\begin{equation}
    \tau_{\rm depl} = \frac{M_{\rm gas}}{{\rm SFR}}.
\end{equation}
We list the gas masses and depletion times for our galaxies in Table \ref{tab:derived_properties}. For GN-CL-3, which has both CO(5--4) and CO(3--2) detections, we show values for CO(5--4) for consistency with GN-CL-2, but the values derived from both lines are consistent within errors\footnote{Specifically, $M_{\rm gas} = (12.3\pm2.3)\times10^{10}~{\rm M_\sun}$ and $\tau_{\rm depl} = 41\pm13~{\rm Myr}$ for CO(5--4), and $M_{\rm gas} = (12.8\pm2.7)\times10^{10}~{\rm M_\sun}$ and $\tau_{\rm depl} = 43\pm14~{\rm Myr}$ for CO(3--2).}. The depletion times of $\sim$50 Myr are consistent with values for other high-redshift SMGs based on high-$J_{\rm up}$ CO lines \citep[e.g.,][]{Casey16}. Based on the calculations in \citet{Casey16}, we would not be likely to observe two SMGs in the same structure with such short depletion times, though we note that values of $\tau_{\rm depl}$ are dependent on the SFR measurement methods used, and, by extension, the assumed SFH, IMF, and conversion factor, $\alpha$.

\subsection{(Sub)Millimeter Evidence of a Protocluster} \label{sec:submmprotocluster}

As mentioned previously, the three SMGs we targeted with NOEMA are part of a larger grouping of seven bright SCUBA-2 sources to the northwest of the {\em HST\/} coverage of the GOODS-N. One is confirmed by our NOEMA observations to lie at high redshift ($z = 4.42$), while another (source 7 in Figure \ref{fig:sourcemap}) has a $U$-band counterpart and $z_{phot} = 1.58$, making it unlikely to belong to the same $z \approx 3.14$ halo as GN-CL-2 and GN-CL-3. The remaining three sources (4, 5, and 6 in Figure \ref{fig:sourcemap}) have SCUBA-2 850~$\mu$m fluxes of 9.8~mJy, 8.6~mJy, and 8.2~mJy, respectively. NOEMA redshift scans of these additional sources, which will determine whether they are part of the same system as GN-CL-2 and GN-CL-3, will be carried out in 2021. In the meantime, we can use the known number counts of bright SMGs to estimate the probability that these sources belong to a larger structure or protocluster.

The S2COSMOS survey of \citet{Simpson19} presents number counts of SCUBA-2 850 $\mu$m sources in the COSMOS field over 2.6 deg$^2$, with a typical $1\sigma$ noise level of 1.2 mJy beam$^{-1}$ in the central region of the field. Their incompleteness-corrected cumulative number counts (their Table 2) suggest that one can expect 61.9$^{+7.2}_{-7.7}$ sources per deg$^2$ (0.017$^{+0.002}_{-0.002}$ arcmin$^{-2}$) at a flux density $S_{850} > 7.7$ mJy. However, there are seven $S_{850} > 8$ mJy galaxies in the northwest offshoot of the GOODS-N that occupy an area only $\sim4.6 \times 4.1$ arcmin$^2$ in size, for a cumulative source density of $N(S_{850} > 8$ mJy) = 0.371 arcmin$^{-2}$. If we assume that the positions of bright SMGs are completely random on the plane of the sky and uncorrelated in redshift space, then the number counts in \citet{Simpson19} suggest that we should expect an average of just 0.321 sources with $S_{850} > 8$ in a given $4.6 \times 4.1$ arcmin$^2$ area, a factor of $21.8$ lower than what is observed in our field. Even removing sources 1 and 7 (for a total of 5 galaxies across a $4 \times 4$ arcmin$^2$ area) yields a projected overdensity 14 times higher than what would be expected if the remaining sources were not part of a single structure.

Alternatively, let us assume that whether or not a bright SMG is seen in a unit area is a Poissonian process so that we can use small number statistics \citep{Gehrels86} to determine the raw likelihood of seeing multiple randomly-distributed $S_{850} > 8$ mJy sources in an 18.9 arcmin$^2$ box. Specifically, if we treat GN-CL-2 and GN-CL-3 as a single system due to their nearly-identical redshifts, then we may compute the chances of observing $2 < N < 6$ projected systems via $P(N; 0.321) = \frac{\lambda^N e^{-\lambda}}{N!}$, where $\lambda = 0.321$ is the mean expected number of sources in a $4.6\times 4.1$~arcmin$^2$ area. As may be expected, the probability of such a projection drops rapidly with increasing $N$, falling from 3.7\% at $N = 2$ to $1.1\times10^{-4}$\% at $N = 6$. For $N = 5$ (6), only $\sim$2500 (160) such projections are expected to be seen across the entire sky.

However, we know from the combination of our NOEMA observations and the UV through radio photometry of \citet{Hsu19} that sources 1 and 7 are, indeed, a chance projection with one another and with sources 2 and 3. In other words, all three systems are independent ``events" such that $N$ is at least 3. If even one of the remaining three bright SMGs is \textit{also} a chance projection with these systems, then we begin to move into a regime of such extraordinarily low probability that it defies our assumption of independent and physically unassociated systems.

\subsection{Optical Evidence of a Protocluster}

\newcommand{\sigmalow}{$\sigma_{zphot,low}$}
\newcommand{\sigmahigh}{$\sigma_{zphot,high}$}

\begin{figure}[t]
    \centering
    \includegraphics[width=0.9\linewidth,trim={0.25cm 0.1cm 0.2cm 0.1cm},clip]{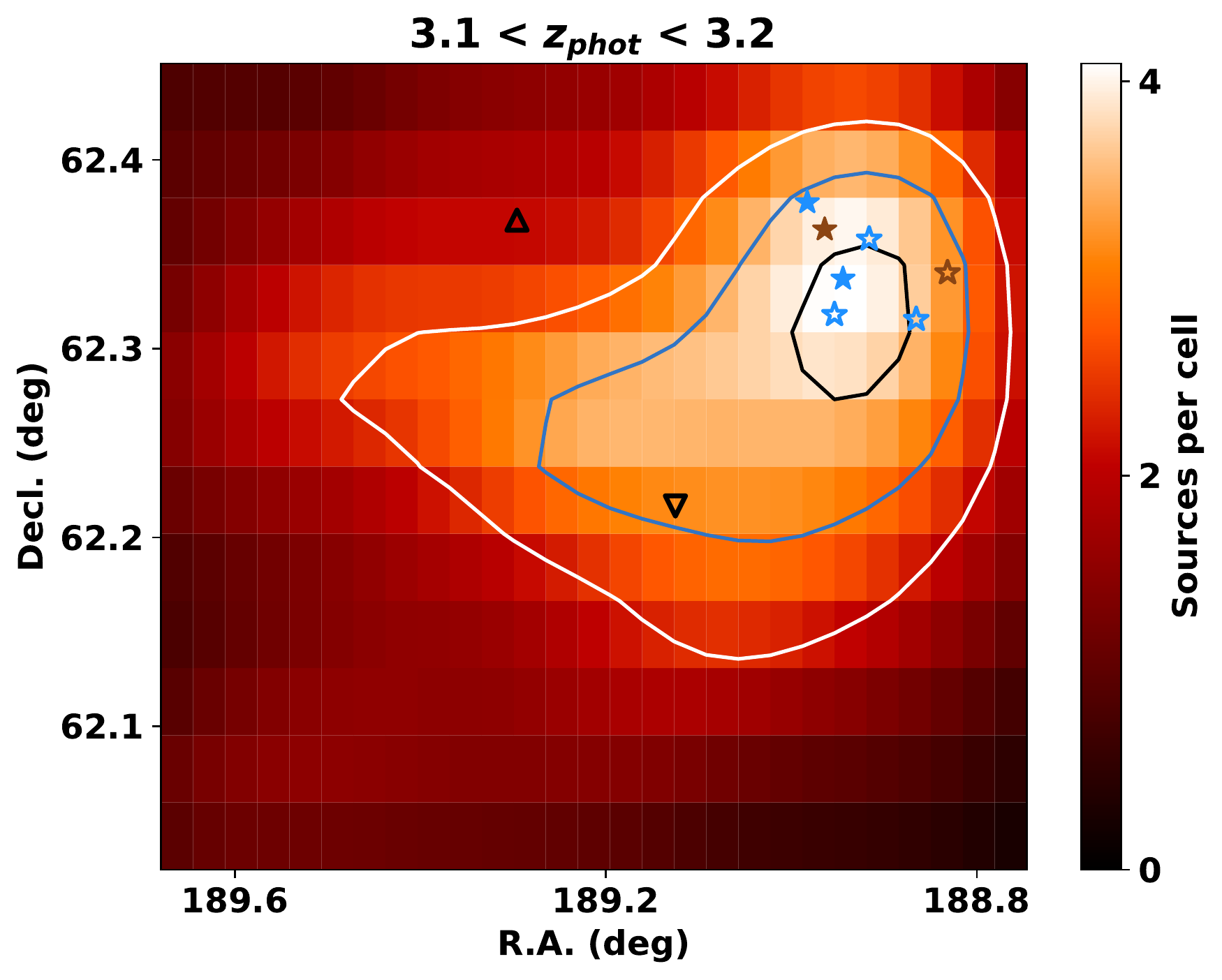}
    \includegraphics[width=0.9\linewidth,trim={0.25cm 0.1cm 0.2cm 0.1cm},clip]{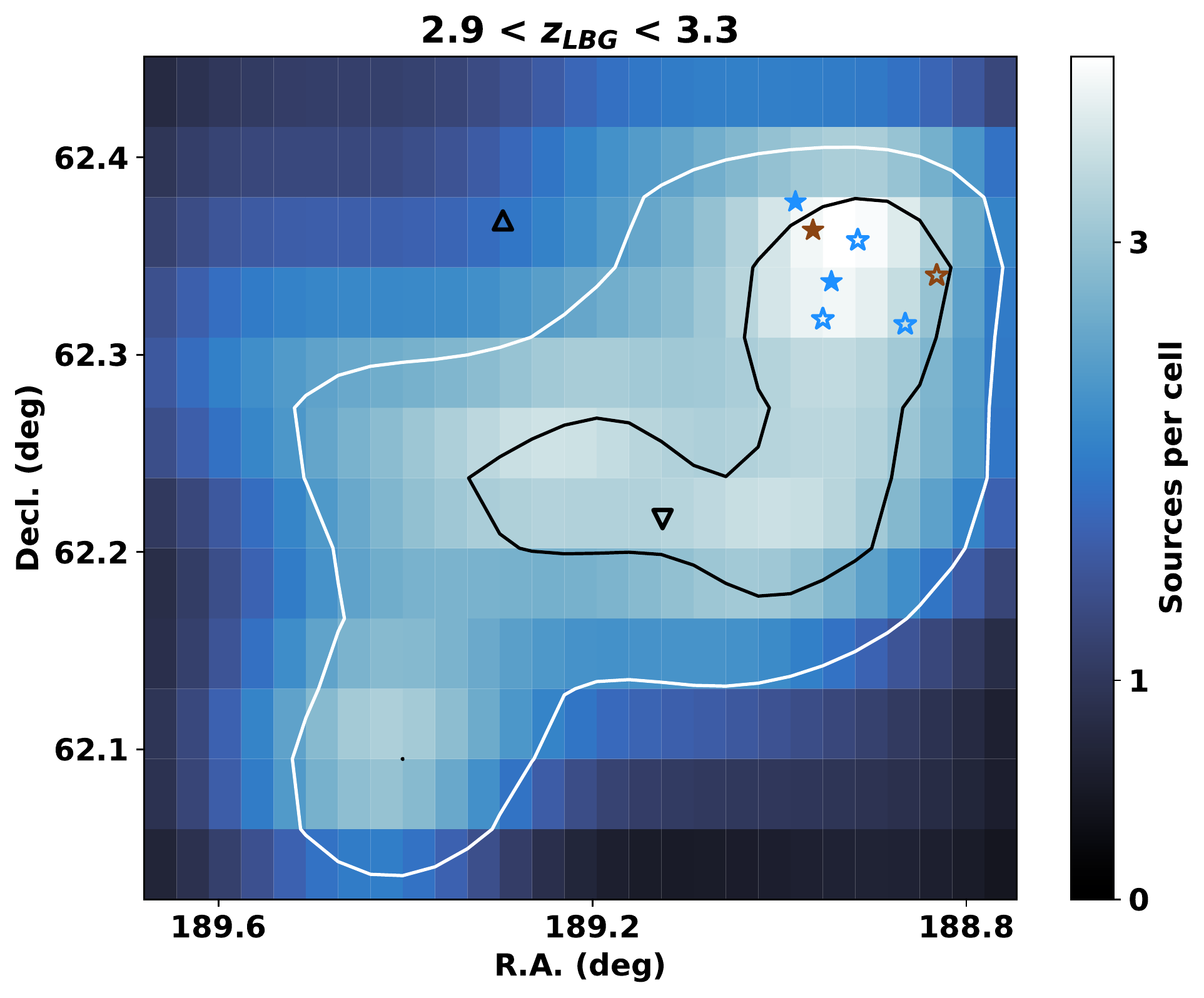}
    \caption{Projected 2D density of all sources in the extended GOODS-N with $3.1 < z_{phot} < 3.2$ in \citet{Hsu19} (\textit{top}) and those in the interval $2.9 < z_{phot} < 3.3$ that meet the LBG criteria of \citet{Capak04} (\textit{bottom}). In both panels, a Gaussian kernel density estimate has been applied, and each cell is $2\times 2$~arcmin$^2$. Contours on the top (bottom) panel indicate regions where the source density is 3$\sigma$, 4$\sigma$, and 5$\sigma$ (2$\sigma$ and 3$\sigma$) higher than the mean in the CANDELS portion of the field. Filled stars show the positions of SMGs observed as part of this work (blue if they are spectroscopically confirmed to lie at $z \sim 3.14$, and brown otherwise). Blue open stars show the positions of other $S_{850} > 8$ mJy SMGs in the region that do not yet have spectroscopic confirmation but may belong to the overdensity. The brown open star is an SMG with a bright $U$-band counterpart and $z_{phot} = 1.58$. Upward- and downward-pointing open triangles, respectively, mark the positions of the GN20 overdensity \citep{Pope05,Daddi09a} and the $z = 1.99$ SMG-rich protocluster  of \citet{Chapman09}.
    }
    \label{fig:Optical}
\end{figure}

The region we consider in this work is well outside the {\em HST\/} coverage of the GOODS-N. Thus, spectroscopic followup is extremely incomplete, even for relatively bright objects, compared to the GOODS/CANDELS portion of the field. Within a 5$'$ radius of the mean position of all seven bright SMGs, there are 316 (2949) sources with $R$ magnitudes brighter than 22 (25), of which only 3 (8) objects have a published spectroscopic redshift in the multiband catalog of \citet{Hsu19}, with none at $z > 2$.

In M. Rosenthal et al., in preparation, we will use our Keck/MOSFIRE spectroscopy of optical-NIR sources in the northwest region of the GOODS-N, together with our existing and upcoming NOEMA data, to characterize the $z \approx 3.14$ protocluster. For now, we use the multiband catalog of \citet{Hsu19} to see whether evidence of an optical overdensity is present based on photometric redshifts only. We restrict to objects with Kron $R$ magnitudes brighter than 25 and $R$ magnitude errors fainter than 26.75, corresponding to a $>5\sigma$ detection.
%, and objects with $3.1 < z_{phot} < 3.2$.
%These limits correspond to a slab of space that is 15.8 proper (91 comoving) Mpc thick. 
We further restrict to regions where the per-pixel $J$-band flux uncertainties are less than twice the median rms in the central part of the field. This results in the loss of 139~arcmin$^2$ of the $JHK_s$ survey area but also removes the bulk of spurious or extremely noisy sources near the edge of the field. 
%We are left with 151 sources across a $\sim 25.4\times 25.4$~arcmin$^2$ area, of which 32 are within $5'$ of the average position of the seven bright SMGs.
Over the redshift interval $3.1 < z_{phot} < 3.2$, we find a mean density of 0.58 galaxies  per arcmin$^2$. However, when we perform a similar analysis in an area of radius 4$'$ centered on the bright SMG overdensity, we find a mean density of  0.89 galaxies per arcmin$^2$. The total number of objects in the overdense region is 45 compared with an expected value of 29.

We next compare the distribution of the number of sources in $1'$ square cells in the full field with that in an $8'$ square area centered on the SMGs (64 cells). A one-tailed Mann-Whitney test of these two samples yields a $p$ value of 0.0014, which implies that they are not drawn from the same distribution and that the optical-NIR overdensity is statistically significant.

For visualization purposes, we show in the top panel of Figure \ref{fig:Optical} a 2D histogram of the source density in this narrow photometric redshift interval. This density map has been ``smoothed" using a Gaussian kernel density estimate with $2' \times 2'$ bins. We note a clear maximum in the density of optical sources that is coincident with the grouping of bright SMGs.

As an additional check, we examine the spatial distribution of candidate Lyman Break Galaxies (LBGs) across the extended GOODS-N in the bottom panel of Figure \ref{fig:Optical}. We use the color criteria of \citet{Capak04} to select LBGs in the redshift interval 
$2.9 < z_{phot} < 3.3$.
% and perform the same Monte Carlo simulations as described above, counting the number of LBGs in 5000 randomly-placed $2' \times 2'$ boxes in the {\em HST\/} portion of the field and another 1000 in the smaller off-center region.
While the photometric redshift interval is too broad to constrain the number of LBGs that may belong to a single, coherent structure at $z \approx 3.14$, we nevertheless note that the maximum density of LBGs is again coincident with the (projected) overdensity of bright SMGs (see the bottom panel of Figure \ref{fig:Optical}). 
%A one-tailed Mann-Whitney test of the two distributions yields a $p$ value of 0.03 ($U$ value 976). Together with the simple calculations presented in Section \ref{sec:submmprotocluster}, these lines of evidence point to the existence of an SMG-rich overdensity or protocluster of galaxies at $z \approx 3.14$.

Finally, we note that the $\sim 0\fdg07 \times 0\fdg07$ angular extent of our extremely bright SMG overdensity is intermediate in size compared to other high-redshift, SMG-rich protoclusters in the literature. For example, this structure is quite compact compared to the $z = 1.99$ structure in the GOODS-N \citep[0\fdg17 $\times$ 0\fdg17]{Chapman09}, the SSA22 protocluster at $z = 3.1$ \citep[0\fdg33 $\times$ 0\fdg50]{Steidel98}, and the $z = 5.18$ HDF850.1 overdensity \citep[0\fdg10 $\times$ 0\fdg13]{Walter12}. However, it is more extended than the GN20 overdensity \citep[0\fdg01 $\times$ 0\fdg01]{Pope05,Daddi09a,Daddi09b} and the Distant Red Core \citep[0\fdg02 $\times$ 0\fdg02]{Oteo18,Ivison20}, both at $z \sim 4$. We defer a fuller discussion of the total SFR, stellar mass, halo mass, and physical extent to M. Rosenthal et al., in preparation.

%By our best estimates, the HyLIRG GN-CL-1 may be an isolated DSFG, but GN-CL-2 and 3 are associated in 3D. A protocluster membership confirmation rate of 2/3 is comparable with other high-redshift systems where multiple candidate members with small projected distances have similar observed properties and photometric redshifts \citep[e.g.][]{Zhou20}.

\section{Summary} \label{sec:summary}

We report the results of a millimeter spectroscopic campaign with NOEMA to measure redshifts for three of the 850~\micron{}-brightest SMGs in the extended GOODS-N, which lie in a close (projected) grouping. We determined unambiguous spectroscopic redshifts for two of our three targets using the NOEMA data alone (GN-CL-1 at $z = 4.42$ and GN-CL-3 at $z = 3.13$) and for the remaining target using both NOEMA and Keck/MOSFIRE spectroscopy (GN-CL-2 at $z = 3.15$). With nearly identical redshifts, GN-CL-2 and 3 are likely part of a single system and may signpost an overdensity of galaxies at $z \approx 3.14$. More importantly, there are three more bright, neighboring SMGs which, based on number counts and simple probability estimates, are extremely likely to belong to the same structure, constituting a protocluster of short-lived SMGs in the high-redshift Universe. Finally, our best-fit SEDs suggest that GN-CL-1, one of the brightest SCUBA-2 850~\micron{} sources in the extended GOODS-N field, is an extremely FIR-luminous, high-redshift dusty starburst with $\sim10^{11}~{\rm M_\sun}$ already formed when the Universe was only 1.4 Gyr old.

\acknowledgments

We thank the anonymous referee for a useful report that helped us to improve the manuscript. We gratefully acknowledge support for this research from a Wisconsin Space Grant Consortium Graduate and Professional Research Fellowship (L.J.), NASA grant NNX17AF45G (L.L.C.), the William F. Vilas Estate (A.J.B.), and the Kellett Mid-Career Award from the University of Wisconsin-Madison Office of the Vice Chancellor for Research and Graduate Education with funding from the Wisconsin Alumni Research Foundation (A.J.B.). 

This work is based on observations carried out under project number W19DG with the IRAM NOEMA Interferometer. IRAM is supported by INSU/CNRS (France), MPG (Germany) and IGN (Spain). L. Jones thanks Vinod Arumugam for many helpful discussions on the reduction and processing of NOEMA observations.


% \vspace{5mm}
\facilities{Subaru (Suprime-Cam), JCMT (SCUBA-2), SMA, NOEMA, {\em Herschel}, JVLA}


\software{astropy \citep{astropy}}

{}



\end{document}